\documentclass[12pt]{article}

\pdfoutput=1

\usepackage{subfigure}
\usepackage{scicite}
\usepackage{graphicx}
\usepackage{times}

\newenvironment{sciabstract}{%
\begin{quote} \bf}
{\end{quote}}

\newcounter{lastnote}
\newenvironment{scilastnote}{%
\setcounter{lastnote}{\value{enumiv}}%
\addtocounter{lastnote}{+1}%
\begin{list}%
{\arabic{lastnote}.}
{\setlength{\leftmargin}{.22in}}
{\setlength{\labelsep}{.5em}}}
{\end{list}}

\title{Frequency metrology in quantum degenerate helium: Direct measurement of the 2~$^3$S$_1$ $\rightarrow$ 2~$^1$S$_0$ transition}

\author
{R. van Rooij$^1$, J.S. Borbely$^1$, J. Simonet$^2$, M.D. Hoogerland$^3$, \\
K.S.E. Eikema$^1$, R.A. Rozendaal$^1$ and W. Vassen$^{1\ast}$ \\
\normalsize{$^1$ LaserLaB Vrije Universiteit, De Boelelaan 1081, 1081 HV
Amsterdam, the Netherlands} \\
\normalsize{$^2$ \'{E}cole Normale Sup\'{e}rieure, Laboratoire Kastler-Brossel,
24 rue Lhomond, 75005 Paris, France} \\
\normalsize{$^3$ Department of Physics, University of Auckland, Private Bag
92019, Auckland, New Zealand} \\
\normalsize{$^\ast$ w.vassen@vu.nl}
}

\date{}

\begin{document} 

\maketitle 

\begin{sciabstract}
Precision spectroscopy of simple atomic systems has refined our understanding of 
the fundamental laws of quantum physics. In particular, helium spectroscopy has 
played a crucial role in describing two-electron interactions, determining the 
fine-structure constant and extracting the size of the helium nucleus. Here we 
present a measurement of the doubly-forbidden 1557-nanometer transition connecting 
the two metastable states of helium (the lowest energy triplet state 2~$^3$S$_1$ 
and first excited singlet state 2~$^1$S$_0$), for which quantum electrodynamic 
and nuclear size effects are very strong. This transition is fourteen orders of 
magnitude weaker than the most predominantly measured transition in helium. 
Ultracold, sub-microkelvin, fermionic $\mathbf{^3}$He and bosonic $\mathbf{^4}$He 
atoms are used to obtain a precision of 8$\times$10$^{\mathbf{-12}}$, providing 
a stringent test of two-electron quantum electrodynamic theory and of nuclear 
few-body theory. 
\end{sciabstract}\pagebreak

\begin{figure}
\includegraphics[width=12 cm]{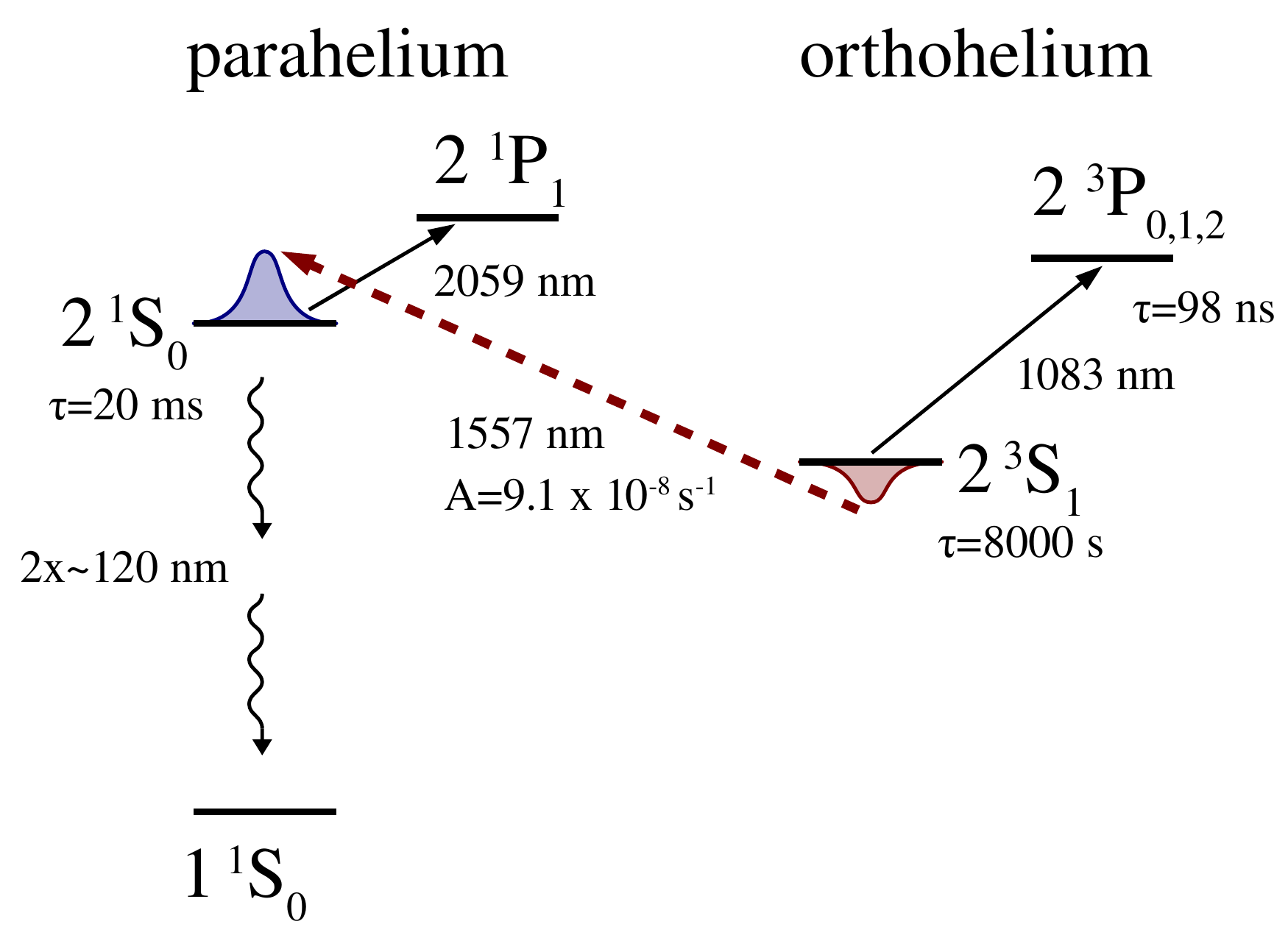}
\caption{\label{fig:scheme} 
Relevant energy levels, transition wavelengths and state lifetimes of helium. 
The magnetic-dipole transition connecting the 2~$^3$S$_1$ state and the 2~$^1$S$_0$ 
state has a wavelength of 1557~nm and an Einstein-A coefficient of 9.1$\times$10$^{-8}$~s$^{-1}$. 
A focused 1557-nm laser also constitutes a trap for ultracold atoms 
in the 2~$^3$S$_1$ state because it is red detuned from the 
2~$^3$S$_1$ $ \rightarrow $ 2~$^3$P$_J$ transitions. As the 1557-nm laser light is 
blue detuned from the 2~$^1$S$_0$~$\rightarrow$~2~$^1$P$_1$ transition, atoms in the 
2~$^1$S$_0$ state are anti-trapped.
}
\end{figure}

The first observations of helium emission spectra at the end of the 19th century 
revealed two separate series of lines, associated with orthohelium and parahelium 
respectively. In 1926 Heisenberg explained the distinction between these two spectra 
on the basis of wave mechanics, electron spin and the Pauli exclusion 
principle~\cite{Heisenberg:26a}. The spectrum of orthohelium arises from triplet 
states for which the electron spins are parallel, whereas in parahelium the electron 
spins are anti-parallel, forming singlet states (Fig.~\ref{fig:scheme}).
From the lowest state of orthohelium, the 1s2s~$^3$S$_1$ state (denoted 2~$^3$S$_1$), only excitations to triplet 
states have been observed. Orthohelium transitions from the 2~$^3$S$_1$ state 
and associated studies of the $n~^3$P$_{0,1,2}$ ($n$\textit{=2,3}) fine-structure 
splittings~\cite{dorrer:97,pastor:04,mueller:05,smiciklas:10,borbely:09,zelevinsky:05} 
have enabled tests of quantum electrodynamics (QED)~\cite{drake:08, yerokhin:10} 
as well as a determination of the fine-structure constant~\cite{pachucki:10, smiciklas:10}. 
In the singlet spectrum of helium (parahelium), electric-dipole transitions from 
the 1~$^1$S$_0$ ground state~\cite{kandula:10} and from the metastable 2~$^1$S$_0$ 
state~\cite{sansonetti:92, lichten:91}
have also provided tests of high-precision QED calculations.  All these frequency 
metrology studies have been performed using either atomic beams or gas discharges. 
However, helium in the metastable 2~$^3$S$_1$ state (He*, lifetime $8\times10^3$~s) can be 
laser cooled and trapped, which allows much longer interaction times for excitation of weak transitions.
He* atoms have been cooled to $\mu$K temperatures, revealing quantum statistical 
effects of bunching and antibunching~\cite{jeltes:07} and achieving quantum degeneracy 
for both the bosonic isotope $^4$He~\cite{robert:01b, pereira:01} 
and the fermionic isotope $^3$He~\cite{mcnamara:06}.

Here we observe an orthohelium -- parahelium transition, 
specifically, the 1557-nm transition between the metastable 2~$^3$S$_1$ 
and 2~$^1$S$_0$ states (Fig.~\ref{fig:scheme}), both in $^4$He and $^3$He. 
This transition is an excellent testing ground for fundamental 
theory of atomic structure. Due to a large electron density at the nucleus, the 
energy of S-states is the most sensitive to QED and to nuclear size effects~\cite{drake:08}. 
For the 2~$^3$S$_1$ and 2~$^1$S$_0$ metastable states, QED terms contribute 4 and 
3 GHz respectively, to a total binding energy of 10$^6$ GHz~\cite{drake:08,yerokhin:10}. 
The present accuracy in the QED calculations is 2~MHz, based on an estimate of 
non-evaluated higher-order terms. 
Many of these terms are common between the isotopes. Therefore, in the calculation of the 
isotope shift (i.e., the difference between the transition frequencies for $^4$He and $^3$He) 
mass-independent terms cancel and the uncertainty is reduced to the sub-kHz level~\cite{drake:05}. 
As the finite nuclear charge radius shifts the 2~$^3$S$_1$ state by 2.6~MHz and the 
2~$^1$S$_0$ state by 2.0~MHz, an accurate isotope-shift measurement allows a 
sensitive determination of the difference in the mean charge radius of the 
$\alpha$-particle and of the $^3$He nucleus, providing a stringent test of nuclear 
charge radius calculations and experiments~\cite{drake:06}.

The natural linewidth of the 2~$^3$S$_1 \rightarrow$ 2~$^1$S$_0$ transition is 8~Hz, 
determined by the 20-ms lifetime of the 2~$^1$S$_0$ state which relaxes via 
two-photon decay to the ground state (Fig.~\ref{fig:scheme}). This transition is
200~000 times narrower than the natural linewidth of the 2~$^3$P state, which is 
most prominently used for spectroscopy in helium. The Einstein $A$-coefficient for 
the 2~$^3$S$_1$$\rightarrow 2~^1$S$_0$ magnetic-dipole transition is 
$\sim$10$^{-7}$~s$^{-1}$~\cite{lach:01, leeuwen:06}, fourteen orders of magnitude 
smaller than for the electric-dipole transition from 2~$^3$S$_1$ to 2~$^3$P$_{0,1,2}$ 
states, which indicates that excitation requires high power and/or long interaction times. 

\begin{figure}
\includegraphics[width=12 cm]{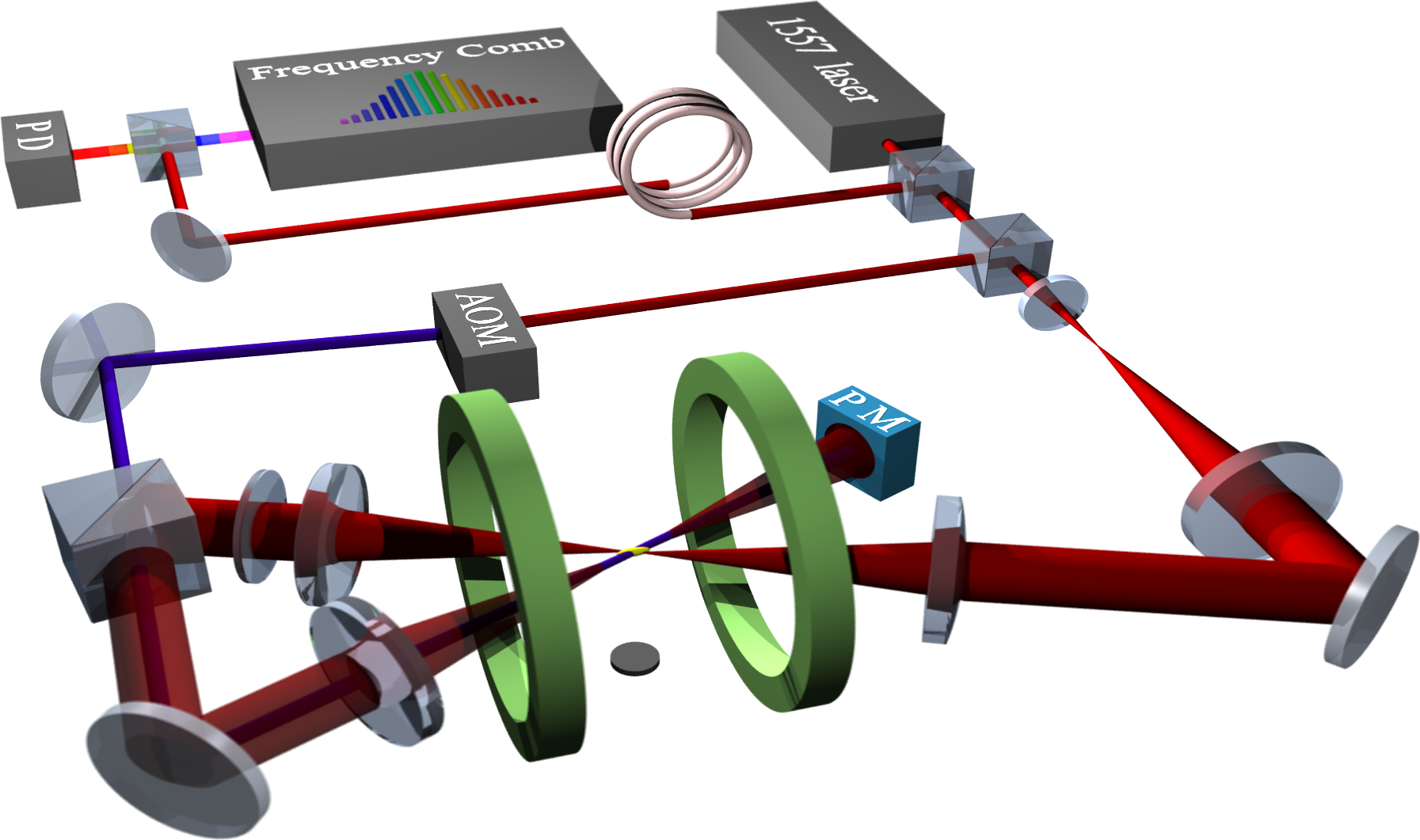}
\caption{\label{fig:setup}
Experimental setup. A small fraction of the 1557-nm laser light is split off and 
coupled via a fiber-optic link to be referenced to a fiber-based frequency comb. 
A heterodyne signal is monitored on a fast photodiode (PD) to determine the absolute 
frequency of the 1557-nm laser. The remaining light is divided into the trap beam and 
the spectroscopy beam. A crossed-beam dipole trap configuration is realized by 
focusing both the incident and returning trap beam (with orthogonal linear polarizations) 
to a waist of $\sim$85$~\mu$m at the center of the magnetic trap (represented by 
the green coils) under a relative angle of 19 degrees, trapping atoms at the 
intersection. The spectroscopy beam is frequency shifted by a 40-MHz acousto-optical 
modulator (AOM), overlapped with the returning trap beam and absorbed by a thermopile 
power meter (PM). A microchannel plate detector is positioned underneath the trap 
for temperature and atom number determination.
}
\end{figure}

The experiment described here was performed using an apparatus designed for the 
production of quantum degenerate gases of helium~\cite{tychkov:06,mcnamara:06}. 
Briefly, the metastable 2~$^3$S$_1$ state is populated by electron impact in an 
electric discharge. The atomic beam is collimated, slowed and trapped using standard 
laser cooling and trapping techniques on the 2~$^3$S$_1 \rightarrow$ 2~$^3$P$_2$ 
transition at 1083~nm. The atoms, optically pumped to $m_J$=+1, are then transferred to a 
Ioffe-Pritchard type magnetic trap. $^4$He* atoms are evaporatively cooled towards 
Bose-Einstein condensation by stimulating radio-frequency (RF) transitions to untrapped states.
For $^3$He* (in the $F$=3/2 hyperfine state), quantum degeneracy is reached by sympathetic cooling with $^4$He*.
Either one, or both, of the two isotopes are transferred into a crossed-beam optical
dipole trap. This trap consists of two focused 1557-nm laser beams, intersecting at their foci as shown in 
Fig.~\ref{fig:setup}. We transfer up to $10^6$~atoms to this optical trap.

After loading the optical trap, the atoms are illuminated by a separate beam for spectroscopy 
which is derived from the same laser as the optical trap beam, but is switched and
frequency shifted by a 40-MHz acousto-optic modulator. 
A heterodyne signal is set up between the 1557-nm laser and a mode of a femtosecond frequency
comb laser to deduce the absolute frequency of the spectroscopy laser.
The frequency comb is based on a mode-locked erbium-doped fiber laser, 
for which both the repetition rate and the carrier-envelope-offset frequency are
referenced to a GPS-controlled Rubidium clock~\cite{science:som}.

\begin{figure}
  \subfigure{\label{fig:MCP_signal}
    \includegraphics[width=0.5\textwidth]{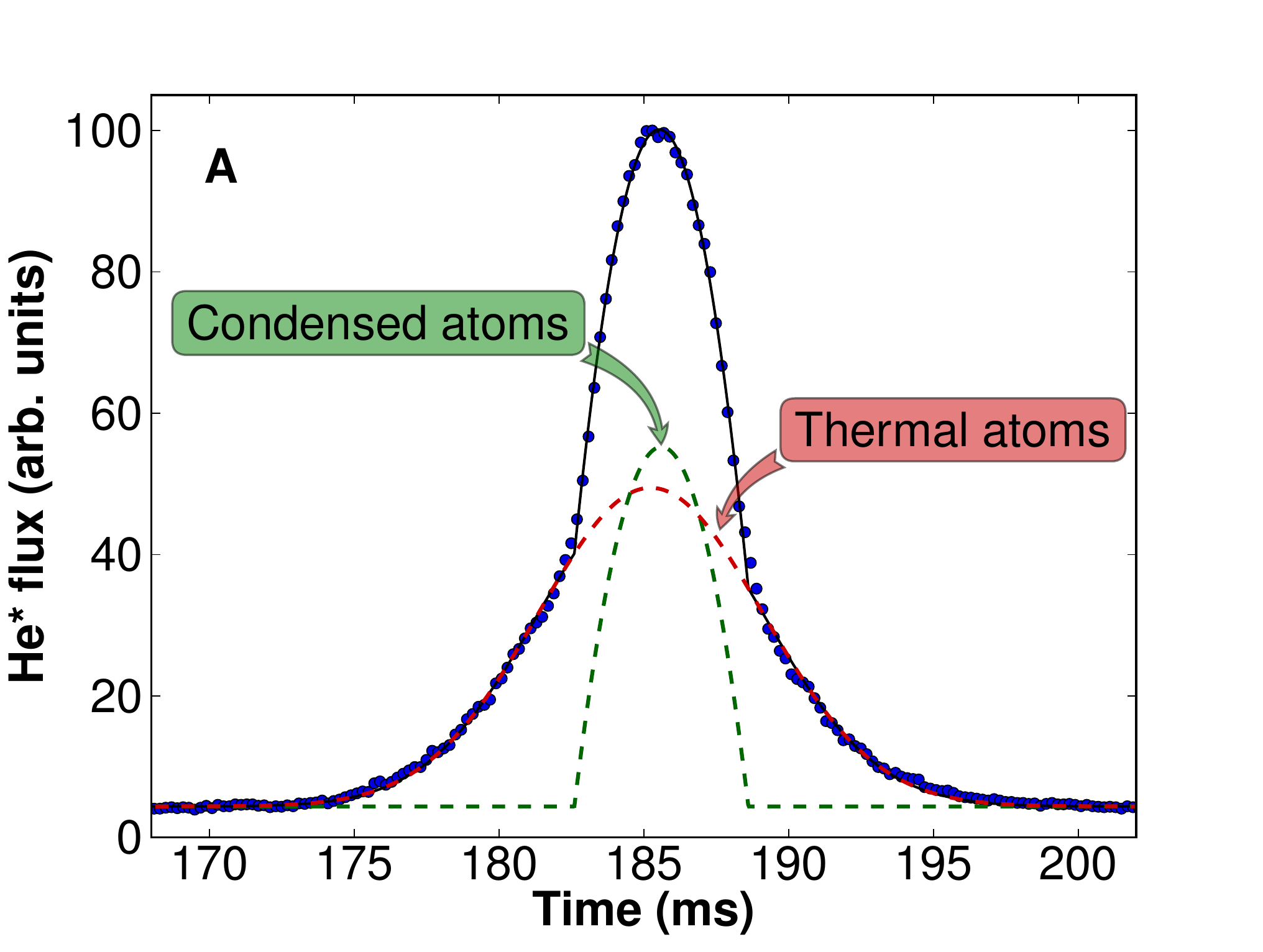}
  }
  \subfigure{\label{fig:resonance}
    \includegraphics[width=0.5\textwidth]{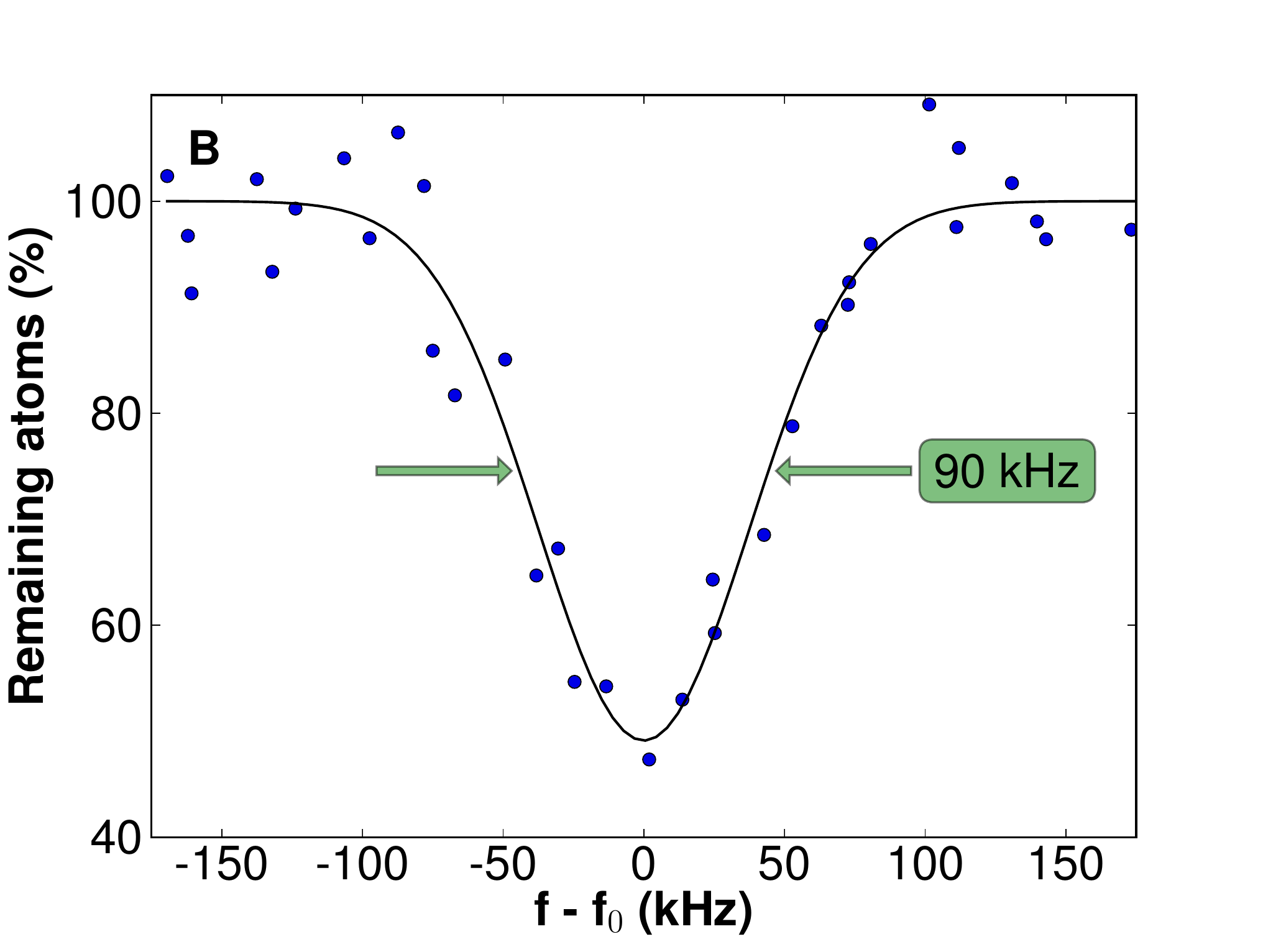}
  }
  \caption{\label{fig:MCP_and_resonance}
(\textbf{A}) Bimodal time-of-flight distribution observed when He* atoms are 
detected on the MCP detector approximately 186 ms after the trapping laser light 
is turned off. The MCP signal is fit to determine the number of Bose-condensed 
atoms. (\textbf{B}) The percentage of Bose-condensed $^4$He atoms 
remaining in the 2~$^3$S$_1$ state as a function of applied laser frequency 
(relative to the fitted center frequency $f_0$). The line is a fit of a Gaussian 
to the data. We measure linewidths varying from 75 to 130~kHz depending on the 
trap depth and on the isotope.
  }
\end{figure}

After a certain interaction time (typically 1 to 6~s), both the 
spectroscopy beam and the trap beam are switched off, allowing the atoms to fall due to gravity.
The high internal energy of He* (20 eV above the 1~$^1$S$_0$ 
ground state) allows for efficient detection on a microchannel plate (MCP) detector
(Fig.~\ref{fig:setup}). The MCP signal reflects both the number of atoms and their temperature. 
In the case of $^4$He, the signal has a bimodal character
that results from the combination of Bose-condensed atoms and thermal atoms (Fig.~\ref{fig:MCP_signal}); a fit to this 
signal provides the number of condensed atoms~\cite{science:som}.
Because the excited state is anti-trapped, the trap is depleted when the 
spectroscopy beam is resonant with the atomic transition. By deducing the remaining number of 2~$^3$S$_1$ atoms
for various laser frequencies, the atomic resonance frequency is determined from a Gaussian fit
to the data (Fig.~\ref{fig:resonance}). The observed linewidth is largely due to the 75-kHz laser linewidth.

Several systematic shifts in the transition frequency are taken into account~\cite{science:som}. The
largest shift is due to the Zeeman effect. The measured transition, 2~$^3$S$_1$~($m_J$=+1) $\rightarrow$ 2~$^1$S$_0$~($m_J$=0)
for $^4$He, and 2~$^3$S$_1$, $F$=3/2 ($m_F$=+3/2) $\rightarrow$ 2~$^1$S$_0$, $F$=1/2 ($m_F$=+1/2)
for $^3$He, is shifted from resonance predominantly by
the Earth's magnetic field. The size of the shift is deduced by measuring the resonance 
frequency of RF spin-flip transitions between the 2~$^3$S$_1$ magnetic substates. 
An additional shift is caused by the momentum transfer from a 1557-nm photon to
an atom. In the case of $^4$He, the high density of the condensate could potentially cause a
mean-field shift~\cite{killian:98}. However, by performing the experiment with 
reduced atomic density, no shift is observed.

\begin{figure}
  \subfigure{\label{fig:result_he4}
    \includegraphics[width=0.5\textwidth]{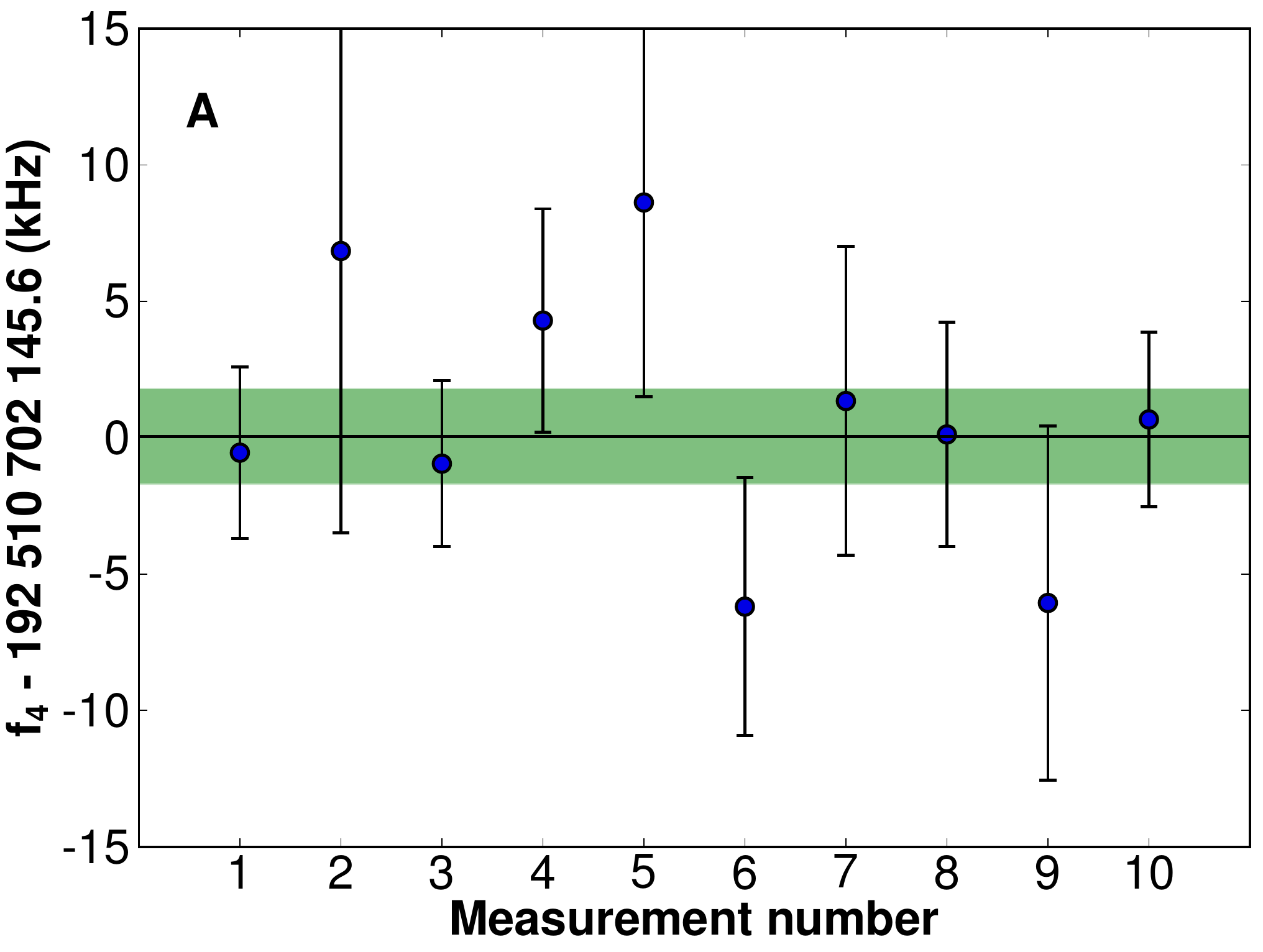}
  }
  \subfigure{\label{fig:result_he3}
    \includegraphics[width=0.5\textwidth]{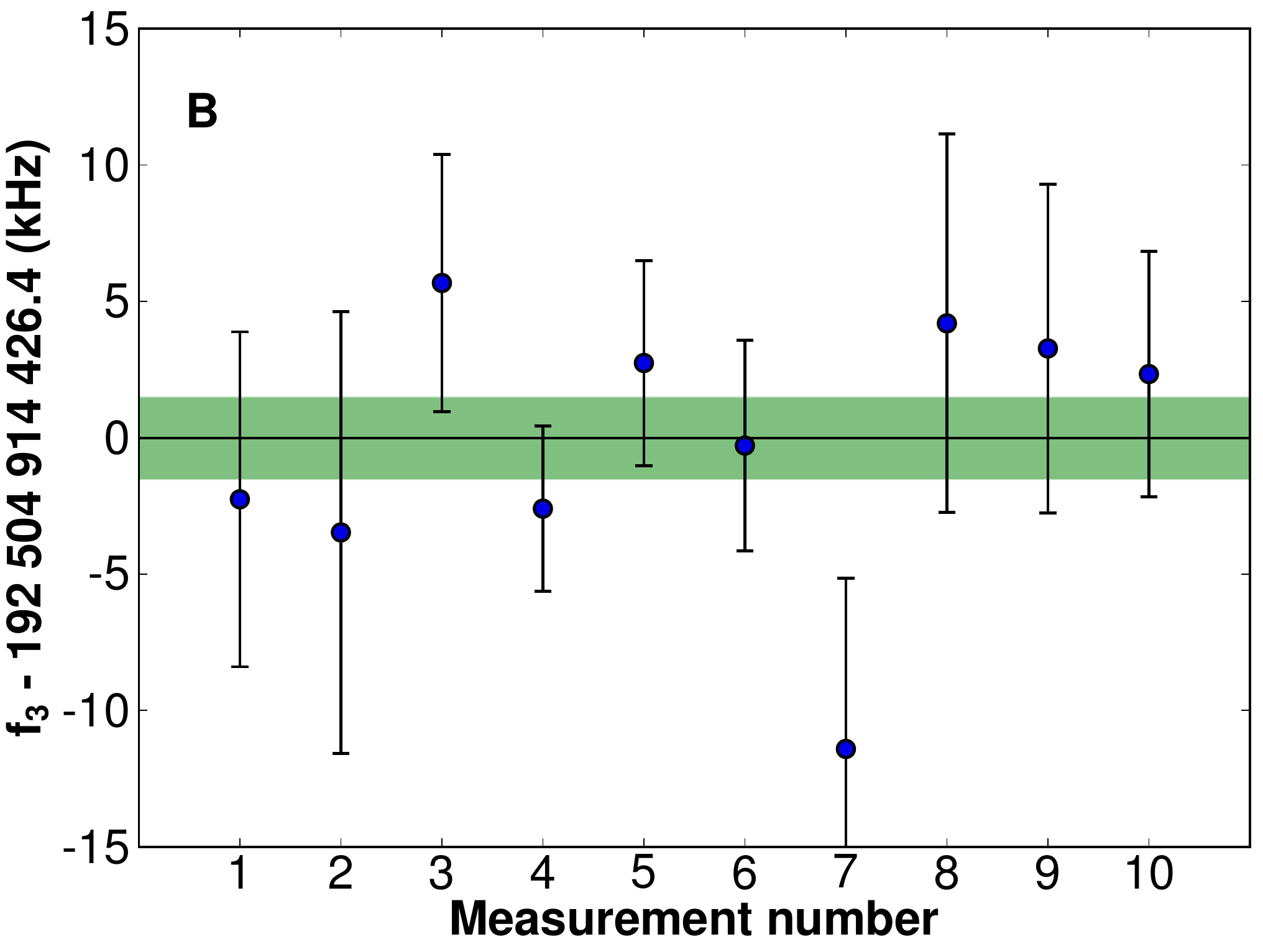}
  }
  \caption{\label{fig:result}
  Measured transition frequencies for $^4$He (\textbf{A}) and for $^3$He (\textbf{B}). 
  The error bar on each data point includes contributions from various systematics, 
  e.g. AC Stark shift and Zeeman shift~\cite{science:som}. The frequencies quoted on 
  the y-axes are the weighted averages of the data points. Their associated uncertainties
  (with the uncertainties due to the frequency comb calibration and the mean field shift 
  added in quadrature) are $\pm$1.8 kHz for $^4$He and $\pm$1.5 kHz for $^3$He, 
  represented by the green bands.
  }
\end{figure}

The second largest systematic frequency perturbation is due to the AC Stark shift 
associated with the intense 1557-nm light which induces the dipole trap:
the specific energy state of the trapping potential for an atom 
determines the AC Stark shift for that atom.
For $^4$He, only excitations of atoms condensed in the ground state of 
the dipole trap are taken into account in determining the transition frequency.
As the trap depth depends linearly on laser intensity, measuring the resonance 
frequency for a range of applied laser powers allows an extrapolation to zero 
laser intensity.
In contrast, $^3$He atoms, due to their fermionic nature, are distributed throughout 
the energy states of the dipole trap and as a result the measured AC Stark shift does not 
equal the trap depth (as is the case with $^4$He), but is reduced due to the density 
of states within the dipole trap. A non-linear shift can then potentially arise 
at high laser intensities, where the larger trap depths allow for higher temperatures. 
To minimize this effect $^3$He atoms are sympathetically cooled to the quantum 
degenerate regime to predominantly populate the lowest energy states of the trapping potential. 
Over the course of several months, 20 independent extrapolations were obtained (as shown in Fig.~\ref{fig:result}) to
deduce an absolute frequency of the 2~$^3$S$_1$~$\rightarrow$~2~$^1$S$_0$ transition 
for $^4$He of $f_4=192~510~702~145.6(1.8)$~kHz and for $^3$He ($F$=3/2 $\rightarrow$ $F$=1/2) 
of $f_3=192~504~914~426.4(1.5)$~kHz, where the one standard deviation error in parentheses 
includes all statistical and systematic uncertainties.

For both isotopes, our result agrees with QED calculations of the ionization energies 
of the two metastable states~\cite{morton:06, yerokhin:10}. The present experimental error in the 
transition frequency is three orders of magnitude smaller than estimates of 
non-evaluated higher-order terms in state-of-the-art QED calculations and 
presents a significant challenge for groups involved in atomic structure theory.

An indirect value of the energy difference between the 2~$^3$S$_1$ and the 2~$^1$S$_0$ 
states can be obtained from the literature (only for $^4$He) by combining experimental transition 
frequencies from both metastable states to high-lying S, P and D states with 
theoretical values for the ionization energies of these states. 
This procedure yields ionization energies for the 2~$^1$S$_0$ state~\cite{lichten:91, sansonetti:92, drake:08} 
and the 2~$^3$S$_1$ state~\cite{dorrer:97, drake:08} and the difference between these 
values gives a transition frequency of 192~510~701.96(16)~MHz, in agreement with 
our result though with hundred-fold lower precision.

Isotope shift measurements, combined with high-precision QED theory, provide a 
method to isolate contributions due to finite nuclear size effects. The difference 
in nuclear charge radii between $^3$He and $^4$He is determined by comparing experiment 
and theory. Because the $^4$He nuclear charge radius is the most precisely known radius 
of all nuclei determined from electron scattering experiments~\cite{sick:08a}, 1.681(4)~fm, 
a value of the $^3$He nuclear charge radius with similar precision can be deduced.
In calculating the isotope shift, QED theory is more precise than our measurement 
as mass-independent terms cancel. 
The theoretical value for the isotope shift (assuming point-like nuclei) is 
8~034~148.6(7)~kHz~\cite{science:som}. Subtracting the measured transition frequencies 
and correcting for the accurately-known hyperfine structure~\cite{morton:06,rosner:70}, 
we find an isotope shift of $f_4$ - $f_3$ + $f_{hfs}$ = 8~034~367.2(2.3)~kHz.
The 218.6~kHz difference may be attributed to the finite size of both nuclei. This nuclear shift is 
proportional to the difference in the nuclear charge radii squared,
$\Delta r_c^2 \equiv r_c^2 \left( ^3\textrm{He} \right) - r_c^2 \left( ^4\textrm{He} \right)$.
Using the theoretical proportionality constant of 4.6642 fm$^2/$MHz~\cite{drake:05} for the 
measured transition, we deduce $\Delta r_c^2 = 1.019(11)~\textrm{fm}^2$.
$\Delta r_c^2$ represents a more universal parameter than the 
value of the isotope shift as it is obtained from various branches of 
physics. Besides through spectroscopic means, it can be determined from nuclear 
theory and from electron-scattering experiments. Nuclear few-body theory provides
$\Delta r_c^2 = 1.16 \pm 0.12$ $\textrm{fm}^2$~\cite{kievsky:08,drake:05, science:som}, whereas from electron-scattering experiments
$\Delta r_c^2 = 1.01 \pm 0.13$ $\textrm{fm}^2$~\cite{sick:08a,sick:08b}. Comparing the values of $\Delta r_c^2$, we find our
result to be in good agreement but an order of magnitude more precise. 
An independent spectroscopic measurement in helium on the 2~$^3$S$_1$~$\rightarrow$~2~$^3$P$_0$ 
transition~\cite{shiner:95} gives $\Delta r_c^2 = 1.059(3)~\textrm{fm}^2$, obtained 
using the most recent QED calculations~\cite{drake:06}.
Although the measurement precision of the isotope shift for this transition is comparable to our precision, the smaller 
uncertainty in $\Delta r_c^2$ is due to a larger sensitivity to differential nuclear charge effects.
Presently, the accuracy to which the $^4$He charge radius is known sets a lower 
limit on the uncertainty of the $^3$He charge radius determined from helium spectroscopy.
Our measurement presents a value for the $^3$He nuclear charge radius
of 1.961(4)~$\textrm{fm}$.

We have also demonstrated that all of the trapped atoms can be transferred to the 
2~$^1$S$_0$ state, producing a source of ultracold singlet helium. 
Optically trapping these atoms simultaneously with cold 
1~$^1$S$_0$ ground state atoms (produced after two-photon decay) opens up the 
possibility to perform two-photon spectroscopy on the 
2~$^1$S$_0$ $ \leftrightarrow $ 1~$^1$S$_0$ transition~\cite{eyler:08, kandula:10},
where QED and nuclear size effects are strongest.

\pagebreak
\setcounter{page}{1}
\noindent{\bf \large SUPPORTING ONLINE MATERIAL}

\vspace{6mm}
\noindent\textbf{Experimental details}
\vspace{6mm}

\noindent
The 1557-nm laser beam is generated by an NP-Photonics erbium-doped CW fiber laser (Scorpio), 
which has a specified short-term spectral linewidth of $<$~5~kHz 
and has a piezoelectric input for fast frequency tuning. The power used for the optical trap is 
variable and is measured on a thermopile power meter. 
The minimum power required to trap atoms is 15~mW and the maximum used is 250~mW, 
which corresponds to a trap depth ranging from 0.3 to 5~$\mu$K, and axial and 
radial trap frequencies ranging from 
$(\omega_{ax}, \omega_{rad})=2\pi\times(20, 140)$~Hz to $2\pi\times(80, 570)$~Hz.
The lifetime of atoms in the trap is limited by background collisions to approximately 15~s.
After being loaded into the optical trap, the atoms are illuminated by the spectroscopy beam. 
This beam has a waist of 280~$\mu$m at the position of the atoms, much larger than the cloud size. 
A measure for the number of atoms is obtained by letting the atoms fall onto a 
microchannel plate (MCP) detector, located 17~cm below the trap center. The MCP
signal (Fig.~\ref{fig:MCP_signal}) shows an integrated time-of-flight trace of the atoms as they hit the detector.
In the case of $^4$He the signal shows a bimodal character, which is fitted with a linear 
combination of a gaussian and a parabolic function, representing the thermal 
atoms and Bose-condensed atoms respectively~\cite{robert:01b,tychkov:06}.
For $^3$He, a Fermi-Dirac time-of-flight function is used to fit the MCP signal~\cite{mcnamara:06}.

\vspace{6mm}
\noindent\textbf{Systematic shifts and uncertainties}
\vspace{6mm}

\noindent
To obtain the value of the transition frequency several systematic 
shifts need to be evaluated, i.e., the AC Stark shift, the Zeeman shift, the recoil 
shift and, in case of a Bose-Einstein condensate a mean-field shift. A summary of the systematic uncertainties
are shown in Table~S1.
The systematic shifts and uncertainties are discussed in detail below.\\

\begin{table}[h]
\centering
Table S1: Combined systematical and statistical uncertainties. Units are kHz.
\begin{tabular}{lcc}
\hline\hline
\multicolumn{3}{c}{} \\[-11pt]          
        & \hspace{10pt} $^4$He \hspace{10pt} & $^3$He \\ \hline
\multicolumn{3}{c}{} \\[-11pt]
Frequency comb           & 0.4    & 0.4 \\
Zeeman shift             & 0.5    & 0.5 \\
Mean-field shift         & 1.1    & \multicolumn{1}{c}{} \\
Recoil shift             & 0.0    & 0.0 \\
AC Stark shift extrapolation   & 1.4    & 1.4 \\ \hline
\multicolumn{3}{c}{} \\[-11pt]
Total (in quadrature)    & 1.8    & 1.5 \\ \hline\hline
\end{tabular}
\end{table}

\noindent\textit{Frequency comb.}
The 1557-nm laser frequency is locked to a frequency comb based on a mode-locked 
erbium-doped fiber laser (Menlo Systems), for which both the repetition rate and 
the carrier-envelope-offset frequency are referenced to a GPS-controlled Rubidium 
clock (PRS10 Stanford Research). The frequency comb has a long-term frequency 
stability of better than one part in 10$^{12}$. The beat frequency between the 1557-nm 
laser and one of the comb modes is measured on a fast photo-diode. 
The $n^{th}$ comb mode producing the lowest frequency beat with the 1557-nm laser 
is the one for which $n$-times the repetition rate ($f_{rep} \approx 250$~MHz) 
of the comb laser plus the carrier-envelope-offset frequency ($f_{ceo} = 20$~MHz) 
matches the 1557-nm laser frequency most closely, then
$f_{1557~nm}=f_{ceo} + n  f_{rep} \pm f_{beat}.$
In order to determine the absolute frequency of the spectroscopy laser the specific 
comb-mode number $n$ is determined using an accurately calibrated wave meter. 
The beat frequency of the 1557-nm laser with the frequency comb is continuously 
measured and regulated by a feed-back loop to match a set-point frequency. 
The integration time is 30~ms, which is a trade-off between the averaging time 
of the comb laser and the short-term stability of the spectroscopy laser.
This laser locking scheme results in a long-term laser linewidth of 75~kHz (FWHM).
For each instance of measuring the 2~$^3$S$_1 \rightarrow$ 2~$^1$S$_0$ transition 
frequency 40 different values of $f_{beat}$ are chosen, where the interaction time 
between the spectroscopy laser and the atoms is, on average, 3 seconds. This
corresponds to a total interaction time of 120 seconds which results in an expected 
uncertainty of the calibration with the frequency comb of less than 0.4~kHz.\\

\noindent\textit{Zeeman shift.}
The transition frequency is shifted by the differential Zeeman shift of the 
initial and the final state in a residual magnetic field, the magnitude of 
which is approximately 0.5~G, predominantly due to Earth's magnetic field.
In a field this size, only first-order Zeeman effects are of relevance at the precision of our experiment.
$^4$He atoms are initially in the 2~$^3$S$_1$ ($m_J$=+1) state, introducing a 
shift rate of $g_{s} \mu_{B}/h = 2.8025$~MHz/G, where $h$ 
is the Planck constant, $g_s$ is the electron g-factor and $\mu_{B}$ is the Bohr magneton. 
The final state is 2~$^1$S$_0$ ($m_J$=0), which does not 
shift due to the absence of a magnetic moment.
In the case of $^3$He, both the initial and final states have an additional nuclear 
spin contribution. The initial 2~$^3$S$_1$, $F$=3/2 ($m_F$=+3/2) state shifts by
$(g_{s} \mu_{B} - \frac{1}{2} g_{i} \mu_{N})/h$, where $g_{i}$ is the nuclear 
g-factor and $\mu_{N}$ is the nuclear magneton. The final state, 
2~$^1$S$_0$, $F$=1/2 ($m_F$=+1/2), shifts by $- \frac{1}{2} g_{i} \mu_{N}/h$. 
As such, the differential shift rate is $g_{s} \mu_{B}/h$, equal to that of $^4$He.

For both $^4$He and $^3$He, the Zeeman shift is determined by inducing radio-frequency (RF) 
transitions to the $m_J$=0 and $m_J$=-1 magnetic substates in $^4$He (Fig.~S1). 
After a 40-$\mu$s RF pulse at a set frequency, the atoms are released from the dipole trap 
and the spin-state populations are separated during time-of-flight in an applied magnetic 
field gradient. Subsequently, an absorption image is taken from which the relative 
populations of the magnetic substates are obtained. The Zeeman shift equals the
frequency for which most atoms are transferred. On average, the uncertainty of
the value we deduce for the Zeeman shift using this procedure is 0.5~kHz.

Zeeman shift measurements are performed intermittently with 
2~$^3$S$_1 \rightarrow$ 2~$^1$S$_0$ transition measurements in order to compensate
for slow changes in the magnetic field. The standard deviation of the value of 
the Zeeman shift over the course of a day is, on average, 2~kHz which corresponds
to a stability of the magnetic field of $\pm$0.7~mG.

\begin{figure}
\includegraphics[width=0.9\textwidth]{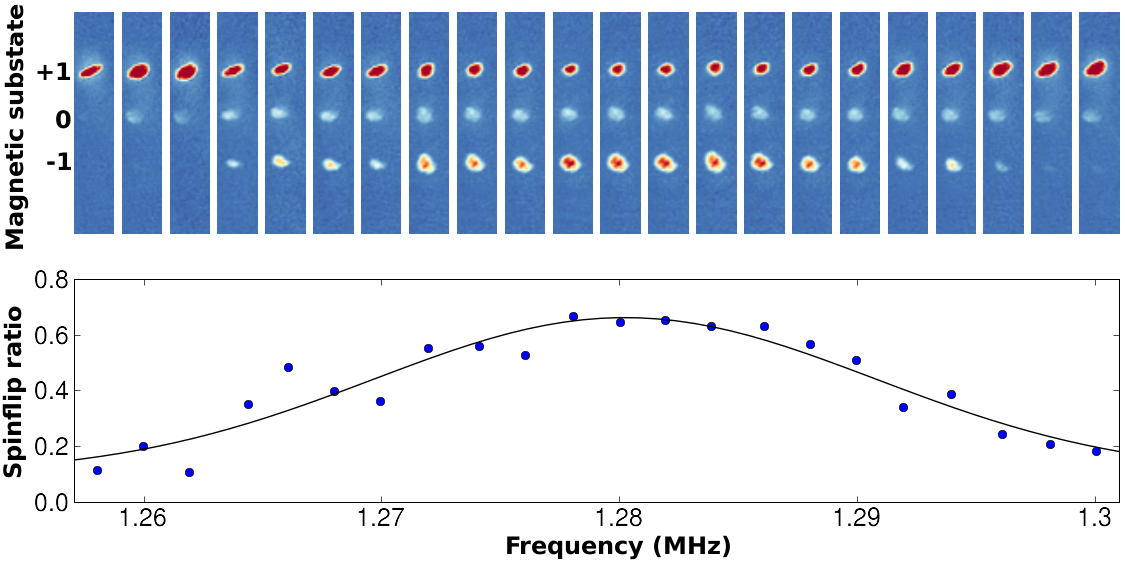}
\\Figure S1: Measurement of the Zeeman shift. The images show the population of magnetic substates 
of the 2~$^3$S$_1$ manifold after a Stern-Gerlach type experiment as a function of 
applied microwave frequency. The ratio of atoms in the $m_J$=0 and $m_J$=-1 state versus 
atoms in the $m_J$=1 state is plotted in the graph. A fit provides the resonance frequency 
which equals the Zeeman shift.
\end{figure}

\noindent\textit{Mean-field shift.}
An additional correction to the transition frequency for $^4$He is the cold collision 
frequency shift~\cite{killian:98}. Originating from the difference in mean-field energy 
between the ground state and the excited state, this frequency shift is proportional 
to the density of trapped atoms and to the difference in $s$-wave scattering lengths 
for the triplet-triplet and the triplet-singlet molecular potentials. The triplet-triplet 
scattering length is +7.5~nm~\cite{moal:06}; however, the value 
of the triplet-singlet scattering length is presently unknown.
By measuring the 2~$^3$S$_1 \rightarrow$ 2~$^1$S$_0$ transition frequency
with a high atomic density and with a low atomic density (reduced by a factor of 5) 
and taking the difference, we find a mean-field shift of $0.07\pm1.08$~kHz which 
is consistent with no mean-field shift to within 1.1~kHz.\\

\noindent\textit{Recoil shift.}
Due to momentum conservation, when an atom is excited from the 2~$^3$S$_1$ state to the 
2~$^1$S$_0$ state, the momentum of the atom is increased by the momentum of the
absorbed photon: $\Delta p = h f / c$, where $h$ is the Planck constant, $f$ is
the photon frequency and $c$ is the speed of light. The corresponding increase
in kinetic energy for the atom must come from the photon, leading to a recoil shift
of $\Delta E = \frac{1}{2 m}\left(\frac{h f}{c}\right)^2$, 
where $m$ is the atomic mass. For a $^4$He atom excited by a 1557-nm photon, the 
recoil shift is 20.6~kHz and for $^3$He it is 27.3~kHz, with negligible uncertainty.\\

\noindent\textit{AC Stark shift extrapolation.}
The AC Stark shift is linearly proportional to the sum of the power in the trap 
and spectroscopy beams. The trap-to-spectroscopy power ratio is 
fixed, but the total power is variable and measured on a thermopile power meter.
The atomic transition is measured several times at a single power and then repeated 
for a range of powers, as shown in Fig.~S2
for $^4$He. From a linear fit, the extrapolation to a
field-free resonance frequency is deduced. On average, one extrapolation gives 
an uncertainty of 5~kHz. In total, 10 extrapolations are determined for $^4$He
and for $^3$He, reducing the final uncertainties to 1.3~kHz.

The thermopile power meter readings are corrected for non-linearity by applying
a bench-top calibration procedure. The uncertainty in this calibration procedure 
contributes an additional uncertainty of 0.6~kHz, leading to a combined uncertainty in 
the AC Stark shift extrapolation of 1.4~kHz.

\begin{figure}
\includegraphics[width=9 cm]{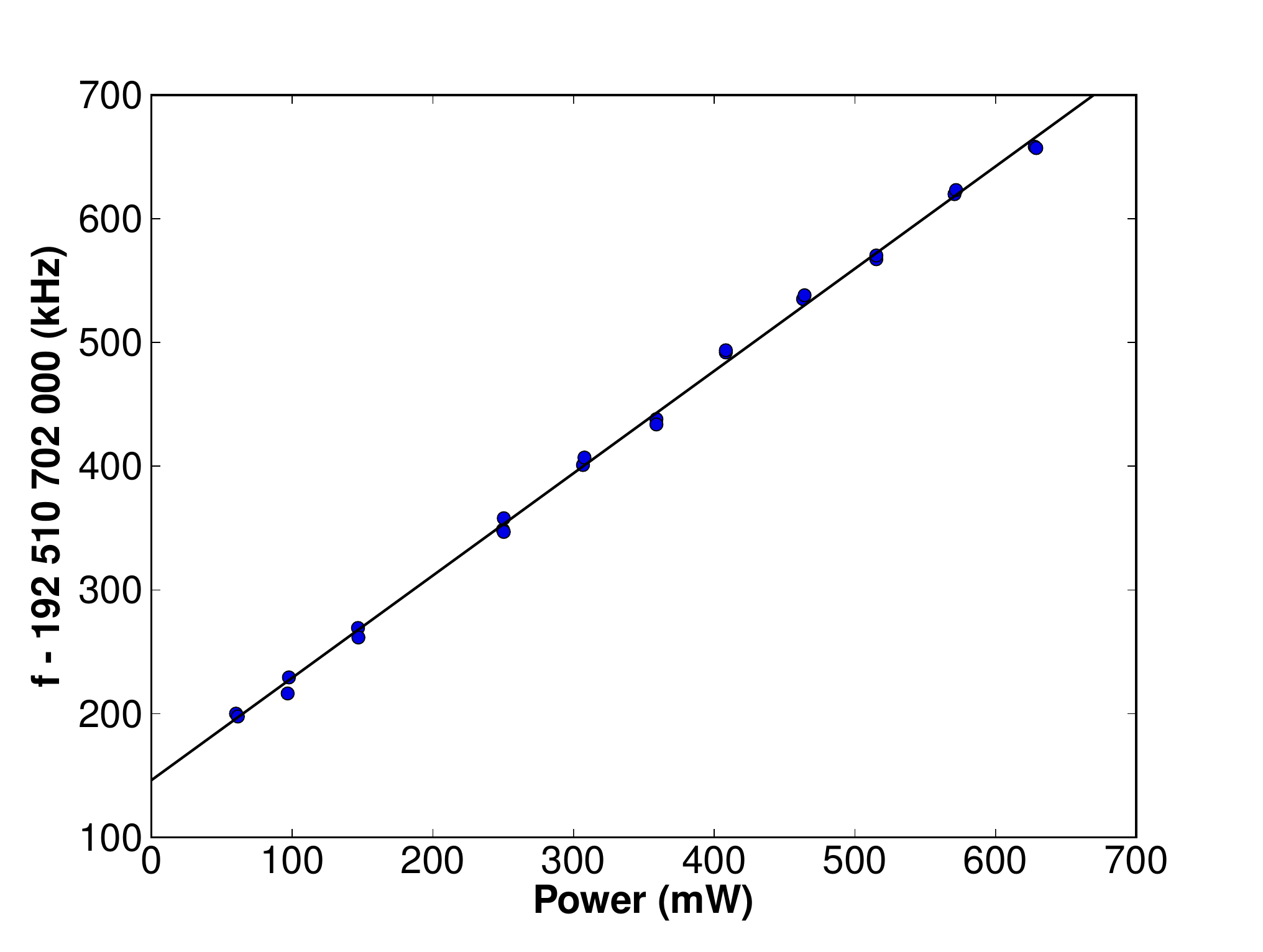}
\\Figure S2: Measured transition frequency for $^4$He as a function of total laser power. The 
resonance frequency is AC-Stark shifted as power increases. The field-free transition 
frequency is found at the point where the fitted line crosses the vertical axis. 
Each data point is obtained by performing several measurements and corrected for 
the Zeeman shift and for the recoil shift.
\end{figure}

\vspace{6mm}
\noindent\textbf{Theoretical determination of ionization energies}
\vspace{6mm}

\noindent\textit{Indirect determination.}
Combining experimental transition frequencies from both metastable states to 
higher S, P and D states with theoretical values for the ionization energies of 
these states yields ionization energies 960~332~041.01(15)~MHz for the 2~$^1$S$_0$ 
state~\cite{lichten:91, sansonetti:92, drake:08} and 1~152~842~742.97(6)~MHz for 
the 2~$^3$S$_1$ state~\cite{dorrer:97, drake:08}. Combining these values a 
2~$^3$S$_1 \rightarrow$ 2~$^1$S$_0$ transition frequency of 192~510~701.96(16) MHz 
is deduced for $^4$He.\\

\noindent\textit{QED theory.}
QED calculations of ionization energies of the two metastable states of $^4$He 
presently agree between different theory groups~\cite{drake:08, yerokhin:10}, 
although their calculations are not fully independent.
Drake and Yan~\cite{drake:08} find 960~332~037.9(1.7) MHz and 1~152~842~741.3(2.5) MHz 
for the ionization energies of 2~$^1$S$_0$ and 2~$^3$S$_1$ respectively, whereas 
Yerokhin and Pachucki~\cite{yerokhin:10} find 960~332~038.0(1.9) MHz and 1~152~842~741.4(2.6) MHz. 
This implies a theoretical transition frequency of 192~510~703.4(2.6) MHz (for both theory groups), 
where we took the maximum of the theoretical uncertainties of both ionization energies 
as the uncertainty in the transition frequency since both uncertainties are expected 
to be correlated (private communication, K. Pachucki and V. Yerokhin, 2010). The uncertainty is due 
to approximations made in calculating all higher-order $\alpha^7$ terms in QED, 
where $\alpha$ is the fine-structure constant.\\

\noindent\textit{Isotope shift.}
From Table 2 of Ref.~\cite{drake:05}, the differences in ionization energies
between $^3$He and $^4$He are 45~862~986.0(5) kHz for the 2~$^1$S$_0$ state and
53~897~130.1(6) kHz for the 2~$^3$S$_1$ state. Recently, Drake corrected a sign 
error in this calculation leading to corrected values of 45~862~980.9(7) kHz and 
53~897~129.4(8) kHz respectively (private communication Drake 2010).
These values agree with the isotope shifts of the ionization energies as 
calculated by Pachucki and Yerokhin (private communication, 2010): 
45~862~979.58 kHz and 53~897~128.33 kHz respectively.
The latter values were calculated applying the methodology of Ref.~\cite{yerokhin:10}.
All these values exclude hyperfine structure effects and contributions due to finite 
nuclear size effects, i.e. assuming point charge radii.
Taking the difference between the isotope shifts of the 2~$^1$S$_0$ and 2~$^3$S$_1$ ionization energies,
the calculated isotope shift for the atomic transition by Drake is 8~034~148.5(7) kHz 
and by Yerokhin and Pachucki is 8~034~148.75(69) kHz, where the uncertainty is estimated to be about 20\% of the E61 term 
which is a sum of pure recoil and radiative recoil corrections of $\mathcal{O}$($\alpha^6m^2/M$), 
where $m$ is the electron mass and $M$ is the nuclear mass. For our calculation of
the nuclear shift we use the average of the theoretical isotope shifts: 8~034~148.63(70) kHz. 
The hyperfine structure of $^3$He~\cite{morton:06}
shifts the 2~$^3$S$_1$ $F$=3/2 state by \mbox{-2~246~587.3} kHz and the 2~$^1$S$_0$ $F$=1/2 state by 
60.6 kHz (due to hyperfine-induced singlet-triplet mixing). 

\vspace{6mm}
\noindent\textbf{Helium charge radii from nuclear theory}
\vspace{6mm}

\noindent Equation~29 from Ref.~\cite{drake:05} relates the nuclear charge radius to an effective
rms radius of the nucleus (corresponding to the distribution of point-like protons and neutrons in the nucleus), to the mean-charge 
radius of the proton and to the mean-charge radius of the neutron. The effective rms radius 
is taken from Tables 7 and 8 of Ref.~\cite{kievsky:08}. The results given in these Tables 
correspond to state-of-the-art calculations of nucleon-nucleon and three-nucleon forces.

\pagebreak

\bibliography{reference2}
\bibliographystyle{Science}

\begin{scilastnote}
\item This work, as part of the European Science Foundation EuroQUAM Programme, 
was financially supported by the Dutch Foundation for Fundamental Research on Matter (FOM).
J.S. acknowledges financial support from the EC's Seventh Framework Programme (LASERLAB-EUROPE). 
M.D.H. and K.S.E.E acknowledge financial support from the Netherlands Organisation for Scientific Research (NWO).
We would like to thank J. Bouma for technical support, and J. Koelemeij and T. van Leeuwen for
fruitful discussions. We also thank G. Drake, K. Pachucki, and V. Yerokhin for sharing the calculated theoretical 
parameters detailed in SOM.
\end{scilastnote}

\end{document}